\documentclass[12pt,a4,epsfig]{article}
\usepackage{amssymb,epsfig}

\newcommand{\beq}[1]{
\marginpar{\small\textsf{#1}}
\begin{equation}\label{#1}}
\newcommand{\eeq}{\end{equation}}
\newcommand{\bea}[1]{
\marginpar{\small\textsf{#1}}
\begin{eqnarray}\label{#1}}
\newcommand{\eea}{\end{eqnarray}}
\newcommand{\beqar}[1]{\begin{eqnarray}\label{#1}}
\newcommand{\eeqar}{\end{eqnarray}}

\newcommand{\inclfig}[2]{\mbox{\epsfxsize=#1cm \epsfbox{#2.eps}}}
\begin{document}

\begin{titlepage}

\begin{flushright}
\begin{tabular}{l}
 TPR - 01 - 03\\
\end{tabular}
\end{flushright}
\vspace{1.5cm}

\begin{center}
{\LARGE \bf
Coulomb dissociation of a fast pion into two jets}
\vspace{1cm}

{\sc D.Yu.~Ivanov}${}^{1,2}$
{\sc L.~Szymanowski}${}^{1,3}$
\\[0.5cm]
\vspace*{0.1cm} ${}^1${\it
   Institut f\"ur Theoretische Physik, Universit\"at
   Regensburg, \\ D-93040 Regensburg, Germany
                       } \\[0.2cm]
\vspace*{0.1cm} ${}^2$ {\it
Institute of Mathematics, 630090 Novosibirsk, Russia
                       } \\[0.2cm]
\vspace*{0.1cm} ${}^3$ {\it
 Soltan Institute for Nuclear Studies,
Hoza 69,\\ 00-681 Warsaw, Poland
                       } \\[1.0cm]

\vskip2cm
{\bf Abstract:\\[10pt]} \parbox[t]{\textwidth}{
 We calculate the electromagnetic contribution to
the scattering amplitude of pion diffractive dissociation
into di-jets which is described by  one photon exchange.
The result shows
that  the factorization procedure known for the description  of
exclusive reactions holds also for this quasi-exclusive process.
We find that the longitudinal momentum distribution of di-jets does not
depend on the form of the pion distribution amplitude.
We discuss the magnitude of the cross section. 
}
\vskip1cm
\end{center}

\vspace*{1cm}

\end{titlepage}

{\large \bf 1.~~} We study the diffractive dissociation of a pion on a nucleus
into two jets. In this process 
a highly energetic pion interacts
with a nucleus and produces  
two jets (di-jets) which in the lowest approximation are formed from the
quark-antiquark ($q\;\bar q$) pair.
\begin{equation}
\label{process}
\pi\;A\;\rightarrow\;q\;\bar q\;A 
\end{equation}

Early studies of this process can be found in  Ref. \cite{FMS93}. 
Recently it was measured in the E791 experiment \cite{exp00} 
and was the subject  of further  theoretical investigations
\cite{NSS00,FMS00}. The process is of interest for its
 potential  
to measure  directly  the pion distribution 
amplitude  (i.e. the probability amplitude to find in a pion 
a valence
$q\bar q$ Fock state  with a certain momentum fraction) 
 and to study the effects  of  colour transparency.

The main contribution to the diffractive process (\ref{process})
is due to  Pomeron exchange, or in QCD language, by the colour
singlet gluon ladder. However, 
 this process can also occur as result of the electromagnetic 
interactions between pion and  target nucleus (Coulomb exchange). 
The strenght of the electromagnetic coupling  
is 
 $\alpha\,Z$ ($\alpha$ is the electromagnetic fine coupling constant).
For heavy nuclei $\alpha\;Z$ is not small and therefore one can ask about 
the size
of the 
Coulomb contribution to (\ref{process}). The answer to this question
 is the subject of the present paper.

{\large \bf 2.~~} The cross section for this process is largest  when
 the momentum
transfer to the nucleus is smaller  than the scale of the nucleus form factor
$\Lambda \sim 60\;$MeV (for Platinum, $Z=78$). 
Therefore we consider the process (\ref{process})
in the case that  large 
transverse momenta of the quark jet ($q_{1\perp}$) and  of the antiquark 
one ($q_{2\perp}$) balance each other, 
$|q_{1\perp}+q_{2\perp}| \ll |q_{1\perp}|\,,\;|q_{2\perp}|$. 
The diagrams describing the Coulomb contribution to the process (\ref{process})
 are shown in Fig. 1.
The wavy line denotes  photon exchanged in the t-channel. 
The large
transverse momenta of di-jets result from the hard gluon exchange 
denoted by  the dashed line. 
They supply a hard scale to the process and therefore we neglect the pion
mass.
The incoming pion has momentum  $p_1,\; p_1^2=0$.
The nucleus mass is $M$, its momenta in the final
and in
the initial states are  $p_2$ and $p_3$, respectively.
The large cms energy squared of the process is given by the 
Mandelstam variable
$s=(p_1+p_2)^2$, whereas the small momentum transfer squared  
equal to the photon virtuality $k^2$ is
denoted as
$t=k^2=(p_2-p_3)^2$.


\begin{figure}[h]
\vspace{-0.5cm}
\begin{center}
\mbox{
\begin{picture}(10,6)
\put(0,0){.}
\put(10,0){.}
\put(10,6){.}
\put(0,6){.}
\put(1,0){\inclfig{8}{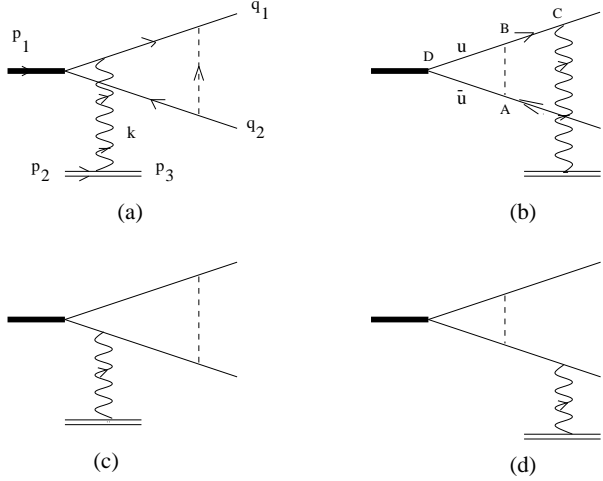}}
\end{picture}
}
\end{center}
\caption{\label{pifig1}
Coulomb contribution  to pion dissociation into two jets.
}
\end{figure}


Let us also introduce  the auxiliary Sudakov vector $p_2^\prime$  
such that $(p_2^\prime)^2=0$
\begin{equation}
\label{p2p}
p_2=p_2^\prime + \frac{M^2}{\bar s}p_1\;,
\end{equation}
where $\bar s =2(p_1p_2^\prime)=s-M^2$.
Then, the Sudakov decompositions of the on-shell 
$q \bar q$
 momenta in the   di-jets are
\begin{equation}
\label{q}
q_1=zp_1+\frac{q_{1\perp}^2}{z\bar s}p_2^\prime + q_{1\perp}
\;\;\;\mbox{and}\;\;\;
q_2=\bar zp_1+\frac{q_{2\perp}^2}{\bar z\bar s}p_2^\prime + q_{2\perp}\;.
\end{equation}
Here, the variable $z$ describes the  fraction of pion 
momentum carried by  quark, $0\leq z \leq 1$.
For the corresponding  fraction of the antiquark momentum 
we use the shorthand notation $\bar z=1-z$.
The decomposition of the photon momentum $k$ reads
\begin{equation}
\label{Sudk}
k= \alpha p_1 + \beta p_2^\prime + k_\perp\,,
\end{equation}
its parameters $\alpha$, $\beta$ and the transverse momentum 
$k_\perp$ are expressed through jets variables
\begin{equation}
\label{albe}
\beta =\frac{M^2_{2j}+(q_{1\perp}+q_{2\perp})^2}{\bar s} \;,\;\;\;
\alpha=-\frac{(q_{1\perp}+q_{2\perp})^2+\frac{M^2\beta}{\bar s}}{\bar s}
\;,\;\;\; 
k_\perp =q_{1\perp}+q_{2\perp} \ ,
\end{equation}
here $M^2_{2j}$ is  the invariant mass of the di-jets 
\begin{equation}\label{invmass}
M^2_{2j}=\frac{q^2_{1\perp}}{z}+\frac{q^2_{2\perp}}{\bar z}-
(q_{1\perp}+q_{2\perp})^2\;.
\end{equation}

As already mentioned the main contribution to the process comes 
from the region of small  momentum transfer (or 
photon virtuality $k^2$) 
$t= k^2=- k_\perp^2+\alpha\beta \bar s $. There
$k_\perp << q_{1\perp}, q_{2\perp}$, which means that 
$q_{2\perp}\approx - q_{1\perp}$ and
\begin{equation}
\label{invmasssmall}
M^2_{2j}=\frac{q^2_{1\perp}}{z\bar z}\;.
\end{equation}
The momentum transfer  $t$ 
 equals $t=k^2 \approx -(k_\perp^2+k_{min}^2)$, and according to
Eq.(\ref{albe}) its minimal value  is
\begin{equation}
\label{kmin}
k_{min}^2= \frac{M^2M_{2j}^4}{\bar s^2}= \frac{M_{2j}^4}{4E^2} \,,
\end{equation}
the last equality being valid in 
 the rest frame of the nucleus in which the pion energy equals $E$.

{\large \bf 3.~~} Our aim is 
to calculate the leading asymptotics of the scattering amplitude
$M$
in powers of $1/q^2_{1\perp} $ (at the leading twist level). 
It is given  as the convolution of the hard scattering
amplitude $T_H(u,\mu_F^2)$ with the pion light-cone
distribution amplitude $\phi_\pi (u,\mu^2_F)$
\begin{equation}
\label{conv}
M=\int\limits_0^1\;du\,\phi_\pi (u,\mu^2_F)\;T_H(u,\mu_F^2)
\end{equation} 
where the hard scattering amplitude $T_H(u,\mu_F^2)$ describes
the production of a free $q \bar q$ pair (the di-jets) in collision of
the $t$-channel photon with  $q$ and $\bar q$
(the later having momenta $u p_1$ and $\bar u p_1$, respectively),
 collinear to the pion momentum
$p_1$.
The factorization scale $\mu_F$ is of the
order of the di-jets transverse momentum $q_{1\perp}$. At lowest order in
 $\alpha_s$, which we consider in this paper, $T_H(u,\mu_F^2)$ does not
depend on $\mu_F$. 

Eq.(\ref{conv})
describes  the factorization procedure in QCD  which disentangles the
contributions to $M$ coming from  large and small distances. 
The soft part  is described 
by the pion  distribution amplitude \cite{BrLe, ChZi}, which is 
defined as follows
\begin{equation}
\label{soft}
\langle 0|\bar d(x) \gamma_\mu 
\gamma_5 u(-x)|\pi^+(p)\rangle_{x^2\to 0}=ip_\mu f_\pi
\int\limits^1_0 du e^{i(2u-1)(xp)} \phi_\pi (u) \ . 
\end{equation}
The constant $f_\pi=131 MeV$ is known experimentally from $\pi\to \mu \nu$
decay, \\ 
$\langle 0|\bar d(0) \gamma_\mu \gamma_5 u(0)|\pi^+(p) \rangle=ip_\mu f_\pi$.

The hard part, i.e. the amplitude $T_H(u)$ in Eq.(\ref{conv}),
 is calculable in the
perturbative QCD, it is given by four tree diagrams shown in Fig. 1.

The factorization procedure described by Eq. (\ref{conv}) is similar 
to that  used for 
 the hard exclusive processes \cite{BrLe, ChZi}.
Its validity in the case of our quasi-exclusive process can be clarified as
follows. 
 Let us consider, for example, the diagram  Fig.1(b). 
The large transverse momentum of quark jets flows along the lines 
 $A\; -\; B\;-\;C$. Therefore their virtualities are much larger than those
of  the other quark lines $D\;-\; A$ and $D\;-\; B$.
Thus, at  leading twist, 
the quark lines $D\;-\; A$ and $D\;-\; B$ have to  be
considered as being on 
mass shell and this part of the diagram
can be factorized out of the hard part
given by the
highly virtual quark and gluon propagators.

Calculating 
the hard amplitude $T_H(u)$ in the Feynman gauge we perfom 
the usual substitution in 
the nominator of the 
photon propagator
\begin{equation}
\label{g}
g^{\mu \nu}\to 2\frac{p_1^\mu p_2^{\prime \nu}}{\bar s},
\end{equation}
where $p_2^{\prime \nu}$ acts on the upper quark vertex.
This brings an inaccuracy $\sim 1/{\bar s}$ which at high energies is
very small.

Since the energy of a photon $k$ is small the
nucleus can be considered as a "scalar particle"
with the electromagnetic formfactor $F_{QED}(k^2)$,
so the photon-nucleus vertex leads to  the factor (see Eq.(\ref{g}))
\begin{equation}
\label{nuclvetrex}
(ieZ)(p_2+p_3)_\mu e^{\mu}F_{QED}(k^2)=(ieZ)(p_2+p_3)_\mu
\frac{p_1^{\mu}}{\bar s}F_{QED}(k^2)=(ieZ)F_{QED}(k^2)\;.
\end{equation}

After calculating convolutions over colour and Dirac indices 
we obtain the contribution of diagram (b) in the form

\begin{equation}
\label{diagrb}
M_{(b)}=-i\,\int\limits^{1}_{0} du \phi_\pi (u)
\frac{(t^at^a)_{ij}}{4N_c}
\frac{2\bar s g^2(e_qeZ) f_{\pi}}{k^2}
\frac{\bar u(q_1) \frac{\hat p_2^\prime}{\bar s}(\hat q_1-\hat k)\gamma_\mu
\gamma_5\hat p_1 \gamma_\mu v(q_2)}{(q_1-k)^2(p_1\bar u -q_2)^2}F_{QED}(k^2)\;.
\end{equation}
The produced quark and antiquark which have the
 colours $i,j$  are described by the 
spinors $\bar u(q_1), v(q_2)$. $g$ is the strong coupling constant.
The electric charge of the nucleus we denote as $eZ$, whereas
$e_q$ is the electric charge of the quark.

The contributions of the other diagrams in Fig. 1 can be writen down
in a similar way. One remark is now in order.
As we stressed above the
hard scattering
amplitude can be calculated
in leading twist accuracy 
 by considering the pion splitting into on-shell
quarks. Then,  the sum of diagrams Fig.1(a) and
 Fig.1(b) as well as the sum of diagrams Fig.1(c) and Fig.1(d) are
separately gauge invariant, i.e. both sums vanish when ${\hat
p}_2^{'}$
in the photon vertex is substituted by ${\hat k}$. 
This property together with  Eqs. (\ref{Sudk},\ref{albe}) permits us
to substitute in all quark-photon vertices 
\begin{equation}
\label{p2}
{\hat p}_2^{'} \rightarrow  -\frac{\bar s}{M_{2j}^2 - t}{\hat k}_\perp\;,
\end{equation}
since the term  proportional 
to $p_1$  in Eq.(\ref{Sudk}) gives in our kinematics
a contribution suppressed by power of $\bar s$. The substitution (\ref{p2}),
being the consequence of  gauge invariance, implies  that
the coupling of the photon to the $q \bar q$ pair vanishes
linearly when $k_\perp \to 0$.

The final result 
for the scattering 
amplitude corresponding to the sum of diagrams shown in Fig. 1 
is given by the formula

\begin{eqnarray}
\label{fullamp}
&& M^{\pi^+ +A\rightarrow 2j + A} = -i
\delta_{i j}\,f_\pi\,\frac{2^6}{3^2}\,\pi^2\,\alpha\, Z\,F_{QED}(k^2)\,
\alpha_s(q_{1\perp})\,
\frac{\bar s}{(k_\perp^2 + k_{min}^2)\,(q_{1\perp}^2)^2} \nonumber \\
&& (e_u\,{\bar z} + e_d
\,z)\,
{\bar u}(q_1)\,\gamma^5\,
\left[ z\, {\hat k}_\perp \, {\hat q}_{1\perp} +
{\bar z}\, {\hat q}_{1\perp} 
{\hat k}_\perp \right] \frac{{\hat p}_2'}{\bar s} v(q_2)
\int\limits_0^1\,du\,\frac{\phi_\pi(u)}{u}\;\;.
\end{eqnarray}
where $\alpha_s=\frac{g^2}{4\pi}$, and we used the symmetry property
$ \phi_\pi(u)=\phi_\pi(\bar u)$.
 The behaviour of the pion distribution
amplitude at the end-points is known:
$\phi_\pi(u)\sim u$ for $u\to 0$, and $\phi_\pi(u)\sim \bar u$ for
$\bar u\to 0$ \cite{BrLe,ChZi}.
 Thus the integral over the momentum fraction 
fraction $u$ is well defined what
confirms that factorization holds
for the process  we discuss.
 Eq. (\ref{fullamp}) is the main result of our paper. 

Let us emphasize that the integral 
of $\phi_\pi(u)$ over $u$
in Eq. (\ref{fullamp}) generates 
 only an overall factor.
Therefore the dependence of the amplitude $M$ on $z$ is universal, i.e.
 it doesn't depend on the shape of
the pion distribution amplitude. It is interesting also to note that the
amplitude (\ref{fullamp}) vanishes for $z=e_u/(e_u - e_d)=2/3$.
Due to the opposite signs of the electric charges of the pion constituents,
 $e_u=2e/3, e_d=-e/3$, the contribution of the 
diagrams (a) and (b) in Fig. 1
cancels the one of diagrams (c) and (d).

The expression in the square bracket on the r.h.s. of Eq.(\ref{fullamp})
can be put in the form
\begin{equation}
\label{onlyone}
{\bar u}(q_1)\,\gamma^5\,[\dots ]\frac{{\hat p}_2'}{\bar s} v(q_2)
=\sqrt{u\bar u} \chi_q \left[(k_\perp q_{1\perp})+
i([k_\perp \times q_{1\perp}\right]\vec \sigma)(z-{\bar z}) ]\chi_{\bar q}
\end{equation}
where $\chi_q, \chi_{\bar q}$ are two dimensional spinors 
of quark and antiquark. 
Since the incoming pion fluctuates into a $q\,\bar q$  states 
with  total helicity  
 zero we have in principle to consider  two amplitudes having different 
spin
configurations of quark and antiquark as denoted by $(\uparrow\downarrow)$ 
and $(\downarrow \uparrow)$.
However,  according to Eq.(\ref{onlyone}) 
$M(\uparrow\downarrow)=M(\downarrow \uparrow)^*$, so  
 effectively  we deal here with only one independent helicity amplitude.

The Colomb contribution to the process (\ref{process}) was considered in
Ref. \cite{FMS00}. Our result (\ref{fullamp}) differs from 
the formula (78) of \cite{FMS00}. 
According to the factorization formula (\ref{conv}) 
we express our result in terms of the pion
distribution amplitude $\phi_\pi(u)$.
We predict the universal $z$ dependence of  
the scattering amplitude $M$, independent of the shape of
$\phi_\pi(u)$,
moreover $M$ vanishes for $z=2/3$. Whereas according to Eq. (78) of
\cite{FMS00} the amplitude is proportional to the lowest pion Fock state
wave function, $M \sim \Psi_\pi(z, q_{1\perp})$ (in our notation).
 This disagreement is, in our opinion, related to the 
treatment  of the
hard gluon exchange in \cite{FMS00}: it is not included 
in the hard scattering block as is usually done within 
the QCD factorization approach but
it is attributed to the high transverse momentum tail of $\Psi_\pi(z,
q_{1\perp})$.

{\large \bf 4.~~} Next,  we estimate the cross section.
For simplicity we use the asymptotic 
pion distribution amplitude 
 $\phi_\pi=6\,u\bar u$.
Using the standard rules  and performing the sum over the colours 
and over the helicities of produced quarks we obtain 
\begin{eqnarray}
\label{exactdifcrospion}
&&(q_{1\perp}^2)^4\;\frac{d\,\sigma^{\pi^+ +A \rightarrow
2j+A}}{d^2\,k_\perp\; d^2\,q_{1\perp}\;dz}
=f^2_\pi\,\frac{2^5}{3\pi}\,\alpha^2\,\alpha_s^2(q_{1\perp})\,Z^2\,
(e_u\,{\bar z} + e_d\,z)^2
\nonumber \\
&&\frac{1}{(k_\perp^2 + k_{min}^2
)^2}\;F^2_{QED}(k^2_\perp)
\;\left[ (z - \bar z)^2\left(q_{1\perp}^2\,k_\perp^2 - 
(q_{1\perp} k_\perp)^2 \right)
+(q_{1\perp} k_\perp)^2 \right]\;.
\end{eqnarray}
For the electromagnetic formfactor of a nucleus (Platinum, $Z=78$)  we use
a naive  dipole approximation
\begin{equation}
\label{nucff}
F_{QED}(k^2)= \frac{\Lambda^2}{\Lambda^2+ k^2},
\;\;\;\;\;\Lambda=60 \mbox{MeV}\;.
\end{equation}


\begin{figure}[ht]
\vspace{-0.5cm}
\begin{center}
\mbox{
\begin{picture}(10,6)
\put(0,0){.}
\put(10,0){.}
\put(10,6){.}
\put(0,6){.}
\put(1,0){\inclfig{8}{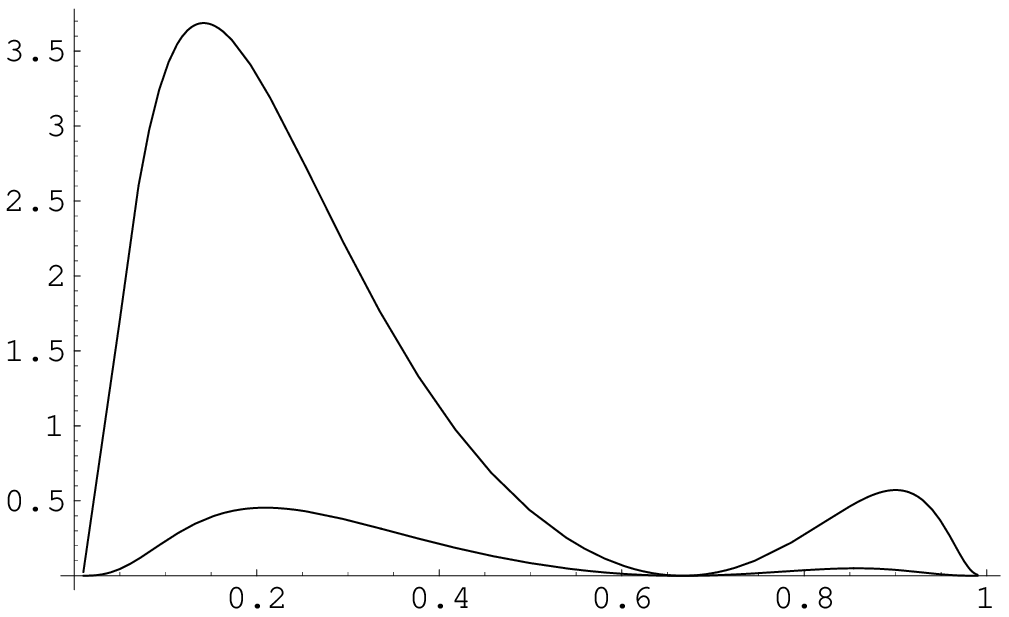}}
\put(0.1,5.3){$q_{1\perp}^6\,\frac{d\sigma}{dq_{1\perp}^2\,dz}\,10^3\;
[\mbox{mbarn GeV$^4$}]$}
\put(9.3,0){$z$}
\put(4,3){$1.25\, \mbox{GeV}$}
\put(2.5,1.0){$2.25\, \mbox{GeV}$}
\end{picture}
}
\end{center}
\caption{\label{pifig2}
Unsymmetrized cross section for pion dissociation into two jets 
on Platinum for
$q_{1\perp}=1.25\,\mbox{GeV}$ and $q_{1\perp}=2.25\,\mbox{GeV}$. 
}
\end{figure}
\noindent In Fig. 2 we present
the differential cross section (\ref{exactdifcrospion}) integrated over
$k_\perp$   for Platinum and the pion energy  $E=500\, \mbox{GeV}$.
Due to the different  charges of $u$ and $d$ quarks the cross section has
an 
asymmetric shape under the 
intechange  $z \leftrightarrow \bar z$. As already
mentioned it is zero for $z=2/3$. 
The cross section vanishes also for $z\to \, 0$ and $z\to\,1$,
this suppression is caused by the electromagnetic formfactor.
According to Eq. (\ref{kmin}), in the vicinity of the end-points
$z(\bar z) \le q_{1\perp}^2/(2E\Lambda)$, the minimal momentum transfer
becomes much larger than
 the scale of the electromagnetic formfactor $\Lambda$.


\begin{figure}[ht]
\vspace{-0.5cm}
\begin{center}
\mbox{
\begin{picture}(10,6)
\put(0,0){.}
\put(10,0){.}
\put(10,6){.}
\put(0,6){.}
\put(1,0){\inclfig{8}{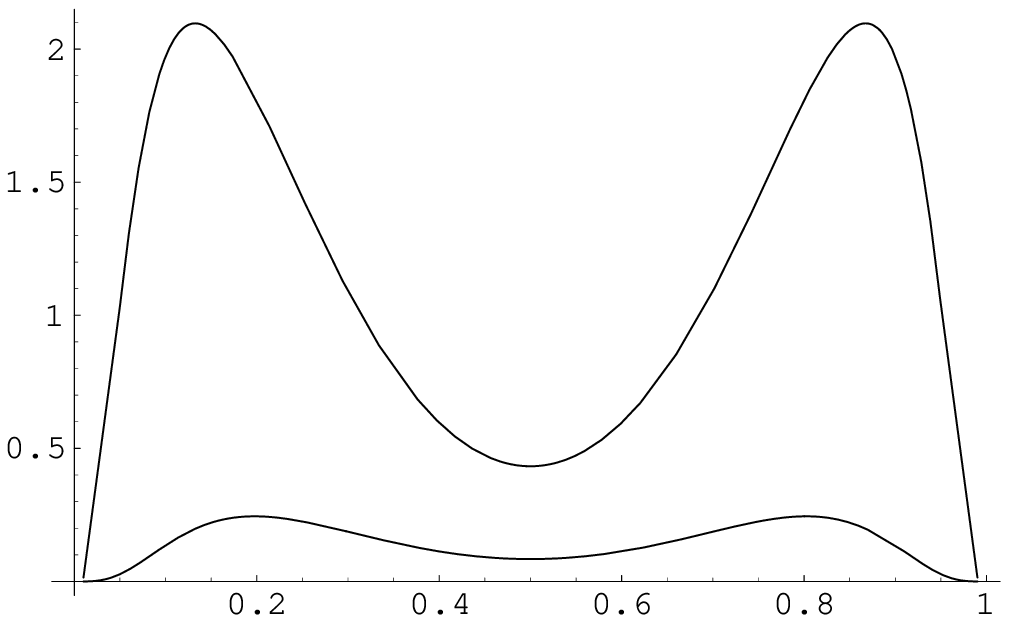}}
\put(0.1,5.3){$q_{1\perp}^6\,\frac{d\sigma}{dq_{1\perp}^2\,dz}\,10^3\;
[\mbox{mbarn GeV$^4$}] $}
\put(9.3,0){$z$}
\put(4,3){$1.25\, \mbox{GeV}$}
\put(2.5,1.0){$2.25\, \mbox{GeV}$}
\end{picture}
}
\end{center}
\caption{\label{pifig3}
Symmetrized cross section for pion dissociation into two jets on Platinum 
for
$q_{1\perp}=1.25\,\mbox{GeV}$ and $q_{1\perp}=2.25\,\mbox{GeV}$.                                     
}
\end{figure}


It is quite difficult to
distinguish jets originating from $u$ and $d$ quarks (by identifying the
leading hadrons), therefore
 we show in Fig. 3
the symmetrized cross section, $d\sigma^{sym}(z)= 1/2(d\sigma (z)
+ d\sigma ({\bar z}))$. The dip structure in the  plot 
for the symmetrized cross
section  is due to above mentioned zero of the amplitude (\ref{fullamp}).  
One could think that the shapes of the plots in Fig. 3 resemble the form of 
the Chernyak-Zhitnitsky  
distribution amplitude \cite{ChZi}, i.e. that 
$d\sigma^{sym} \sim (\phi^{(CZ)}_\pi(z))^2$. This similarity is nevertheless
accidental since
as we explained above the dependence of the amplitude 
$M$ on $z$ is universal
and the shape of curves in Fig. 3 has other reasons.

The Coulomb effect 
studied in this paper is complementary to the contribution
from Pomeron exchange. Although the  QCD part of the   
cross section  is
large  it scales like $1/q_{1\perp}^8$, i.e. it decreases faster at large
$q_{1\perp}$ than  QED contribution which behaves like 
 $\sim 1/q_{1\perp}^6$. 
The relative magnitude of the Coulomb part should also 
be enhanced in the region of
small $z$ or $\bar z$. The 
QCD contribution to the cross section is proportional
to $\phi^2_\pi(z)\cdot F^2_{QED}(t)$ which vanishes 
at $z\; (\bar z) \approx 0$
as $\sim z^2\;(\bar z^2)$.
This behaviour should be compared with the QED contribution
given by
 Eq. (\ref{exactdifcrospion}) which behaves, for  $z\; (\bar z) \approx 0$,
as $\mbox{const}\cdot F^2_{QED}(t)$.


The comparison of our results on the Coulomb diffractive pion dissociation
into two jets with theoretical predictions or with experimental data is 
 difficult.
On the one hand side the theoretical results existing in the literature
\cite{NSS00, FMS00}
for the  QCD contribution are controvercial. In a fortcomming paper
\cite{BISS} we try to solve this issue within the QCD factorization scheme.
On the other hand,  in the kinematical
region covered by  E791 experiment
the Coulomb effect is difficult to observe due to the
 large Pomeron background.
Also, one  cannot compare  our
predictions for Coulomb contribution directly with
 E791 data since their absolute normalization is still
not known.


\vspace*{.5cm}

{\it  \bf Acknowledgments}

\vspace*{.5cm}

\noindent We are grateful to Vladimir Braun, Leonid Frankfurt and Andreas
Sch\"afer for discussions and critical
remarks.

\noindent L.Sz. acknowledges  support by DFG and the warm hospitality at
Regensburg University. 
 D. I. acknowledges the support of Alexander von Humboldt Stiftung.

\end{document}